\documentclass[pre,aps,superscriptaddress,twocolumn,nofootinbib]{revtex4-1}

\usepackage{epsfig}
\usepackage{amsfonts}
\usepackage{latexsym}
\usepackage{amsmath}
\usepackage{amssymb}
\usepackage{slashed}
\usepackage{longtable}
\usepackage{mathrsfs} 
\usepackage[normalem]{ulem}

\usepackage{mathtools}

\usepackage{graphicx}
\usepackage{hyperref}

\numberwithin{equation}{section}

\newcommand{\bea}{\begin{eqnarray}}
\newcommand{\eea}{\end{eqnarray}}
\newcommand{\be}{\begin{equation}}
\newcommand{\ee}{\end{equation}}
\newcommand{\ba}{\begin{align}}
\newcommand{\ea}{\end{align}}
\newcommand{\quotes}[1]{``#1''}

\def\Or[#1]{{\text{O}}\left({#1}\right)}
\def\dotl[#1,#2]{\left\langle #1, #2 \right\rangle}
\def\dotlb[#1,#2]{[ #1, #2 ]}
\def\dotp[#1,#2]{(#1) \cdot (#2)}
\def\aff[#1,#2]{\hat{#1}(#2)}
\def\n4sym{{\cal N}=4 SYM}
\def\>{\rangle}
\def\<{\langle}
\def\weight[#1,#2,#3]{\{(#1),#2,#3\}}
\def\ads[#1]{$\text{AdS}_{#1}$}

  \makeatletter
  \let\over=\@@over \let\overwithdelims=\@@overwithdelims
  \let\atop=\@@atop \let\atopwithdelims=\@@atopwithdelims
  \let\above=\@@above \let\abovewithdelims=\@@abovewithdelims

\usepackage{xcolor}
\definecolor{ogreen} {RGB}{71,191,145}

\begin{document}

\preprint{}

\title{Passive odd viscoelasticity}
\author{Ruben Lier}
\email{rubenl@pks.mpg.de}
\affiliation{Max Planck Institute for the Physics of Complex Systems, 01187 Dresden, Germany}
\affiliation{W\"{u}rzburg-Dresden Cluster of Excellence ct.qmat, 01187 Dresden, Germany}
\author{Jay Armas}
\email{j.armas@uva.nl}
\affiliation{Institute for Theoretical Physics, University of Amsterdam, 1090 GL Amsterdam, The Netherlands}
\affiliation{Dutch Institute for Emergent Phenomena (DIEP), University of Amsterdam, 1090 GL Amsterdam, The Netherlands}
\author{Stefano Bo}
\email{stefabo@pks.mpg.de}
\affiliation{Max Planck Institute for the Physics of Complex Systems, 01187 Dresden, Germany}
\author{Charlie Duclut}
\email{duclut@pks.mpg.de}
\affiliation{Max Planck Institute for the Physics of Complex Systems, 01187 Dresden, Germany}
\author{Frank J\"{u}licher}
\email{julicher@pks.mpg.de}
\affiliation{Max Planck Institute for the Physics of Complex Systems, 01187 Dresden, Germany}
\affiliation{Cluster of Excellence Physics of Life, TU Dresden, 01062 Dresden, Germany}
\author{Piotr Sur\'owka} 
\email{surowka@pks.mpg.de}
\affiliation{Department of Theoretical Physics, Wroc\l{}aw  University  of  Science  and  Technology,  50-370  Wroc\l{}aw,  Poland}
\affiliation{Max Planck Institute for the Physics of Complex Systems, 01187 Dresden, Germany}
\affiliation{Institute for Theoretical Physics, University of Amsterdam, 1090 GL Amsterdam, The Netherlands}
\affiliation{Dutch Institute for Emergent Phenomena (DIEP), University of Amsterdam, 1090 GL Amsterdam, The Netherlands}

\begin{abstract}
Active chiral viscoelastic materials exhibit elastic responses perpendicular to the applied stresses, referred to as odd elasticity. We use a covariant formulation of viscoelasticity combined with an entropy production analysis to show that odd elasticity is not only present in active systems but also in broad classes of passive chiral viscoelastic fluids. In addition, we demonstrate that linear viscoelastic chiral solids do require activity in order to manifest odd elastic responses. In order to model the phenomenon of passive odd viscoelasticity we propose a chiral extension of Jeffreys model. We apply our covariant formalism in order to derive the dispersion relations of hydrodynamic modes and obtain clear imprints of odd viscoelastic behavior.
\end{abstract}

\maketitle

\section{Introduction}

Mechanical responses to the applied external stresses are key in the understanding of materials. Such responses can take a simple elastic form for materials, whose microscopic composition consists of a lattice of atoms, or a more complex viscoelastic behavior for substances with a compound mesoscopic structure. Such media include polymers \cite{ward2004}, metamaterials \cite{Bertoldi2017,shankar2021topological} and biomatter \cite{Kruse2004,Frthauer2012}. Viscoelastic behavior manifests itself through energy dissipation when external stress is applied and then removed. This can be understood as a symptom of a non-vanishing viscosity and thus a fluid-like characterization of a viscoelastic material. As a consequence, a proper description of viscoelastic responses requires a combination of elastic and viscous components in the defining constitutive laws. In general, this is a complicated task with many unknown variables, which requires a phenomenological treatment. The non-equilibrium, dissipative nature of viscoelasticity suggests that it is a transient phenomenon to an equilibrium state that can be either a solid or a fluid. As a result we can consider two distinct classes of materials: viscoelastic solids and viscoelastic fluids. If one assumes that the relation between stress and strain is linear it is possible to write down constitutive relations, for these two classes, parameterized by response coefficients. 

In its simplest incarnation, a linear model for solids is known as the Kelvin-Voigt model and for fluids as the Maxwell model. They are usually visualized in one dimension in terms of connected springs and dashpots. The Kelvin-Voigt solid is modelled by a spring and a damper connected in parallel and the Maxwell fluid is modelled by a spring and a damper connected in series. Although these models are at the core of rheological descriptions of viscoelasticity they are often not capable of capturing experimental stress-strain responses in more complex materials. The Maxwell model does not describe a progressive deformation of a material under constant stress (creep), and the Kelvin-Voigt model does not describe stress relaxation. One way to improve this is to account for higher-order relaxations in the constitutive relations. This can be visualized by extending the basic Kelvin-Voigt and Maxwell models with additional springs and dashpots. Viscoelastic models that consist of two springs and one dashpot are called standard linear solids or Zener models \cite{ward2004} and models built from two dashpots and one spring are called standard linear fluids or Jeffreys models \cite{joseph_fluid_1990}. One-dimensional models can be generalized to higher dimensions, in which the responses are controlled by tensors of coefficients that respect symmetries of a given system. Therefore, for a class of materials that break chirality, new parity and time-reversal symmetry breaking transport coefficients have to be taken into account, even for the simple Kelvin-Voigt and Maxwell models \cite{Banerjee2021}.

Chirality is an asymmetry under mirror imaging. This asymmetry plays an important role in various biological systems \cite{Tsai2005,Henley2009,Vandenberg2009,Henley2012,Frthauer2013,Naganathan2014}, metamaterials \cite{Reinbold2019,shankar2021topological}, condensed matter \cite{son2019chiral,Armitage2018} and high-energy physics \cite{Kharzeev_strongly_2013}. An effective hydrodynamic theory of systems with broken chiral symmetry in two dimensions is distinguished by a new transport coefficient, namely Hall viscosity \cite{Avron1995,Avron1998,levay,Lapa2014,Lucas2014,Kogan,Ganeshan2017,Liao2019,SouslovPRL2019,Soni2019,Markovich2021,Monteiro2021}.

Hall viscosity can exist in a system that breaks parity and time-reversal symmetry and is characterized by a non-dissipative coefficient. It is present, for instance, in various electron
systems in the presence of background magnetic ﬁelds. These systems are passive and break these discrete symmetries by the presence of a magnetic field. Additionally, Hall viscosity can also exist in chirally asymmetric active systems \cite{Soni2019}. Active systems are maintained out of equilibrium by active processes at small scales (e.g. caused by self-propelled or fuel-consuming elements \cite{Julicher_2018,Marchetti2013}). The non-equilibrium state implies that time-reversal symmetry is broken. In contrast, passive systems do not have small scale active processes and relax to a thermodynamic equilibrium in the absence of external forces, which in the absence of a magnetic field typically obey time-reversal invariance.

Chiral active systems can also host odd elasticity. It describes elastic forces that cannot be expressed as a gradient of elastic energy, but stem from a driving out of equilibrium \cite{Scheibner_2020}. Moreover, in the setting of active chiral viscoelastic media, modelled by either the Kelvin-Voigt model or the Maxwell model, it was recently argued that odd material properties are governed by an interplay of Hall
viscosity and an odd elastic coefficient \cite{Banerjee2021}. Signatures of odd elastic and odd viscous properties were recently investigated in experiments with starﬁsh embryos \cite{Tan2021} and magnetic colloids
 \cite{Bililign2021}, respectively. The aim of the present work is to investigate viscoelasticity of chiral systems beyond the simple Kelvin-Voigt and Maxwell models. The main result of our analysis is the phenomenon of passive odd viscoelasticity, where both the viscous and the elastic contributions of the response exhibit odd behavior. In particular, we show that odd viscoelasticity can exist in passive systems with broken time-reversal symmetry. 
Surprisingly, this implies that odd elasticity can transiently exist even in passive systems.

Our study is based on irreversible thermodynamics \cite{Groot1984,landau_fluid_2011, Jensen2012}. In this setting, the existence of an entropy current whose divergence is enforced to be positive semi-definite embodies a local version of the second law of thermodynamics. This imposes non-trivial constraints on the various transport coefficients. This approach can be incorporated in the study of viscoelastic fluids \cite{PhysRevA.6.2401, chaikin2000principles, Armas2020} but traditional studies are based on broken symmetry variables (i.e. Goldstone fields of spontaneous broken translations) and only describe Kelvin-Voigt-type models \cite{Armas2020}. In order to embed more complex rheology models, such as Maxwell and Jeffreys model, into the framework of viscoleastic hydrodynamics, we follow the approach of refs. \cite{Fukuma_2011, Fukuma_2011_Entropic}, in which metric degrees of freedom that describe the evolution of the lattice structure of the material are taken into account. In refs. \cite{Fukuma_2011,Fukuma_2011_Entropic} the case of parity-even systems in three dimensions was studied in detail, in the relativistic context. In this paper, we generalize this approach to chiral systems in two dimensions and extend it to systems with Galilean symmetry. We explicitly demonstrate that parity-odd transport, under appropriate entropic restrictions, can be incorporated into a passive chiral version of Jeffreys model. 

This paper is organised as follows. In sec.~\ref{simplifiedddd} we give a brief overview of our main results and introduce the passive chiral Jeffreys model. In sec.~\ref{sec:oddjeffrey} we provide the details of the model and present the covariant formalism together with the conservation laws and the thermodynamic description of local thermal equilibrium. In sec.~\ref{sec:entropy} we perform a technical analysis of the entropy current and derive the constraints on transport coefficients, constitutive relations and rheology equations from the second law of thermodynamics. In sec.~\ref{modess} we study this model more closely, in particular we look at the modes associated with the odd viscoelastic model. In sec.~\ref{sec:discussion} we discuss some implications of our results and future research avenues.

\section{Summary of results}
\label{simplifiedddd}
The purpose of this paper is to show that, using irreversible thermodynamics, parity-breaking elastic effects can be incorporated into an odd extension of Jeffreys model \cite{jeffrey1,jeffrey2,jeffrey3,jeffrey4,jeffrey5}. The goal of this section is to briefly outline some of the details of this model and the consequences of our analysis. 

Typically, viscoelastic models are defined via phenomenological equations specifying the time evolution of stresses $\tau_{i j }$ in terms of the time evolution of strains $\mathcal{E}_{i j }$, where $i,j=1,2$ are spatial Cartesian indices in two dimensions. We consider the two-dimensional rotationally invariant case in this work. To describe the phenomenological equations for this case we only need four-tensors which satisfy the property
 \begin{align}
 U_{ij k l }    =   U_{j i  k l } =  U_{ij l k  } ~~ .  \label{tensorconstraintttt}
 \end{align}
Isotropic tensors that do not satisfy eq.~\eqref{tensorconstraintttt} correspond to an inception of torque density or a response to rotation \cite{Scheibner_2020,Han2021}. We do not consider such responses in this work, as they are not relevant for the passive viscoelastic systems that we focus on. There are three remaining independent isotropic four-tensors that satisfy eq.~\eqref{tensorconstraintttt} which are given by
\begin{align}
\zeta_{ij k l }  & =  \delta_{ij} \delta_{kl} \label{bulktensorrrr}  \\ 
\eta_{i j k l }    & = \frac{1}{2} \delta_{ k (i} \delta_{j) l}  - \frac{1}{2} \delta_{i j } \delta_{kl }  \label{tracelessss}~~,  \\ 
 \eta_{i j  k l  }^{*  }   &  =  -   \frac{1}{4}  (   \epsilon_{i k } \delta_{j l}  + \epsilon_{i l} \delta_{j k} + \epsilon_{j k } \delta_{i l} +\epsilon_{j l} \delta_{i k}    )~~,  \label{oddtenssorrr}
\end{align}
where $\delta_{kl }$ is the Kronecker delta  and $\epsilon_{k l }$ is the Levi-Civita tensor. We have used a shorthand notation $H_{(ij )  } =   H_{ij   }  + H_{j i}$ to denote the symmetric part of a general tensor $H_{ij}$. Eq.~\eqref{bulktensorrrr} is the trace projector, whereas  eq.~\eqref{tracelessss} introduces the traceless projector while eq. \eqref{oddtenssorrr} is a parity-odd tensor that enables the description of both odd elasticity and odd viscosity in two dimensions. This parity-odd tensor is traceless and satisfies the anti-symmetry property $ \eta_{i j  k l  }^{*  } = -  \eta_{ k l i j   }^{*  } $. In eq.~\eqref{oddtenssorrr} and in the rest of this paper we use the superscript $*$ to denote parity-odd contributions.
 
\begin{figure}[ht]
\includegraphics[width=8cm]{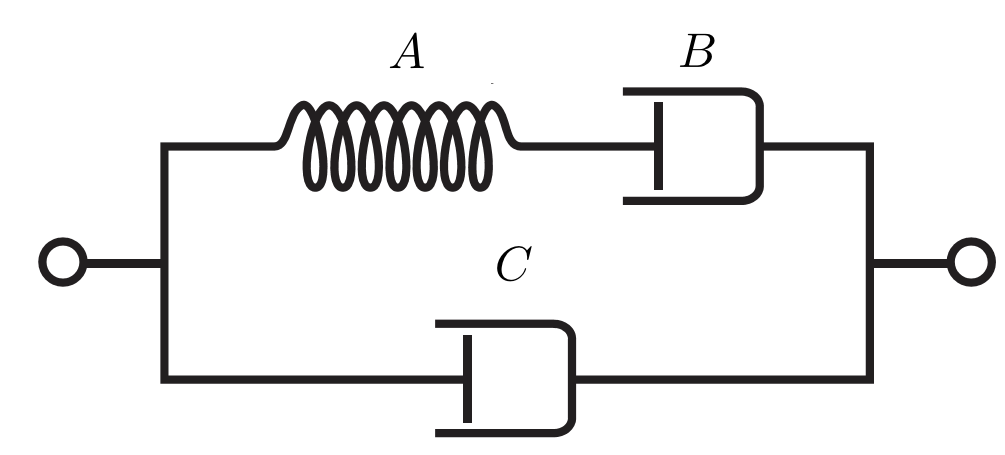}
\caption{Diagram corresponding to the full bulk sector given in sec. \ref{viscoelasticmodelsss}. }
\label{bulksectorsnip}
\end{figure}

\begin{figure}[ht]
\includegraphics[width=8cm]{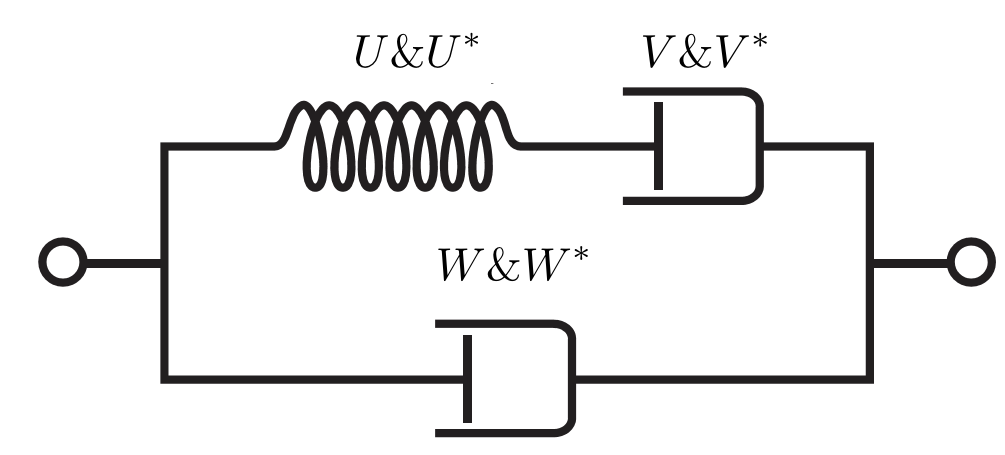}
\caption{Diagram corresponding to the shear sector. \quotes{\&} refers to a parallel connection of an odd and an even component of the same type.}
\label{shearsectorsnip}
\end{figure}

In order to characterise the models that we will soon introduce, it is useful to decompose the stress $\tau_{ij}$, incorporating corrections with respect to local thermal equilibrium into an isotropic stress $\tau$ and a shear stress $\tau_{\langle i j \rangle }$. In particular the stress is given by $\tau_{ij}= \frac{1}{2} \delta_{ij} \tau+\tau_{\langle i j \rangle }$, where we have introduced the short-hand notation $A_{\langle i j \rangle }   = \eta^{\, \, \, k l }_{i j } A_{k l  } $ and $A  =  A_{k   }^{\, \, k }$, holding for any two-tensor $A_{i j }$. Here we used the Einstein summation convention and moreover we have defined the contravariant (upper) index as being raised with $\delta^{ij }$. As in traditional textbooks, it is possible to introduce the phenomenological model of interest, which we refer to as the \emph{odd Jeffreys model}, by means of material diagrams (one for the bulk sector and one for the shear sector). The bulk sector diagram, which does not permit odd contributions, is depicted in Fig.~\ref{bulksectorsnip}. The three-element representation of this constitutive equation, composed of two dashpots and one spring is not unique. In the case of three elements two different configurations corresponding to fluids can be mapped to each other. The nonuniqueness is a general feature that holds in multi-element models.
The bulk Jeffreys model's \cite{jeffrey1,jeffrey2,jeffrey3,jeffrey4,jeffrey5} characteristic equation is given by 
\begin{align}
 \begin{split}
   &    \dot{\tau} +  \sigma \tau    =   - \alpha   \ddot{\mathcal{E} }  -  \tilde \beta     \dot{\mathcal{E}}    \label{haaazzz}~~,
 \end{split}
\end{align}
where we have introduced the notation $\dot{A} \equiv  \frac{\partial}{\partial  t }  A   $ for any tensor object  $A$ with indices supressed. The coefficients in eq.~\eqref{haaazzz} are functions of $A$, $B$, and $C$ introduced in fig.~\ref{bulksectorsnip}. We refer the reader to the app.~\ref{app:diagrams} for the explicit formulae. We note that the bulk sector as determined by eq.~\eqref{haaazzz} does not contain parity-odd terms. However, the shear sector, for which the material diagram is depicted in fig.~\ref{shearsectorsnip} 
is given by the characteristic equation
\begin{align}
 \begin{split}
   &   \dot{\tau}_{\langle  i j  \rangle } 
   +  ( \chi \eta_{i j  }^{  \, \, \, \, \, k l }  + \chi^{*} \eta^{* \, \,  k l }_{i j }   )  \tau_{k l}   \\ &   =  -   ( \gamma \eta_{i j}^{  \, \, \, \, \, k l }  + \gamma^{*} \eta^{*  \, \, k l}_{i j  }   )    \ddot{\mathcal{E}}_{kl}   -  ( \zeta \eta_{i j  }^{  \, \, \, \, \, k l }  + \zeta^{*} \eta^{* \, \, \,  k l }_{i j  }   )    \dot{\mathcal{E}}_{kl}    \label{haahhax}~~,
 \end{split}
\end{align}
contains several parity-odd contributions. Analogous to \eqref{haaazzz}, all transport coefficients in the shear sector are functions of $V$, $V^{*}$, $U$, $U^{*}$, $W$, $W^{*}$ introduced in fig. \ref{shearsectorsnip} (see app.~\ref{app:diagrams} for details). 

The model eqs. \eqref{haaazzz} and \eqref{haahhax} can be obtained from the diagrammatic representations presented in figs.~\ref{bulksectorsnip} and \ref{shearsectorsnip} following analogous rules to those used in the context of electric circuits. In this case the bulk coefficients $\sigma, \alpha, \tilde\beta$ in \eqref{haaazzz} and the shear coefficients $\chi,\chi^*,\gamma,\gamma^*,\zeta,\zeta^*$ are arbitrary. However, the main goal of this paper is to embed these models into a formal hydrodynamic framework of viscoelastic fluids. Within this setup, requiring the second law of thermodynamics to be satisfied leads to nontrivial constraints and relations among the various coefficients.

The fact that the odd Jeffreys model allows us to incorporate odd elastic terms in the sense of refs. \cite{Scheibner_2020,Banerjee2021} is a non-trivial result. In fact, when considering the two models studied in refs. \cite{Scheibner_2020,Banerjee2021}, in particular the \emph{odd Kelvin-Voigt model} given by the model equation 
\begin{align}
 \begin{split}
   &     \tau_{\langle i j \rangle } 
    =    - ( \phi \eta_{i j  }^{  \, \, \, \, \, k l }  + \phi^{*} \eta^{* \, \, \,  k l }_{i j  }   )    \mathcal{E}_{kl}  -  ( \psi \eta_{i j  }^{  \, \, \, \, \, k l }  + \psi^{*} \eta^{* \, \, \,  k l }_{i j  }   )     \dot{\mathcal{E}}_{kl} ~~,   \label{haaaz}
 \end{split}
\end{align}
and the \emph{odd Maxwell model}, which is equivalent to the shear sector of Jeffreys model \eqref{haahhax} but with $\gamma$ and $\gamma^*$ set to zero\footnote{We provide diagrammatic constructions of these models in app.~\ref{app:diagrams}.}
\begin{align}
 \begin{split}
   &   \dot{\tau}_{\langle i j \rangle } 
   +  ( \chi \eta_{i j l }^{  \, \, \, \, \, k l }  + \chi^{*} \eta^{* \, \,  k l }_{i j }   )  \tau_{k l}      = -    ( \zeta \eta_{i j  }^{  \, \, \, \, \, k l }  + \zeta^{*} \eta^{* \, \, \,  k l }_{i j  }   )    \dot{\mathcal{E}}_{kl}    \label{haaahh}~~,
 \end{split}
\end{align}
we find that some parity-odd terms are not allowed within our framework of irreversible thermodynamics. Specifically, the coefficients $\phi^*$ of the odd Kelvin-Voigt model and $\zeta^{*}$ of the odd Maxwell model are required to vanish, leaving behind only non-zero parity-odd contributions due to odd viscosity. Therefore, the two simpler models introduced in refs. \cite{Scheibner_2020,Banerjee2021} can only describe active systems and not passive ones.

\section{The odd Jeffreys model} \label{sec:oddjeffrey}
In this section we elucidate the necessary details of the odd Jeffreys model discussed in the previous section. In particular, we embed the model in a simple hydrodynamic framework and give spatially covariant expressions for the characteristic equations and conservation laws.

\subsection{Strain}
\label{strainnnn}
\begin{figure}
\includegraphics[width=8cm]{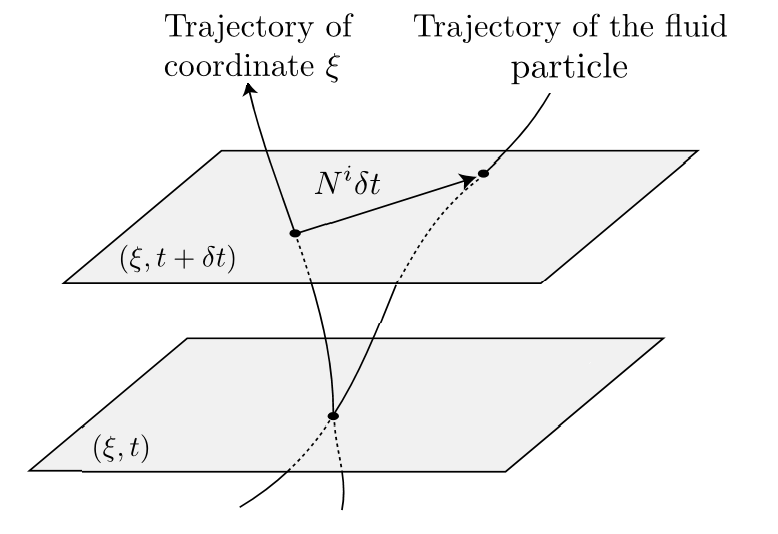}
\caption{The vector $N^i$ describes the motion of the fluid particles with respect to the coordinate frame. }
\label{fehiufehiuh}
\end{figure}

In sec. \ref{simplifiedddd} we introduced several models which depended on the strain $\mathcal{E}_{ij}$. We will now use concepts from differential geometry and a covariant representation to define such strains. This approach will be convenient to introduce plastic strains in the next section via a dynamical intrinsic metric. To this end we consider a material element with label $\xi^i$ which at time $t$ is found in real space at position $X^a $ \cite{Azeyanagi2009}. This defines a function $X^a (\xi^i,t )$. For instance, the label $\xi^i$ can be defined as the position $X^a $ at an initial reference time point. The physical space is endowed with the Euclidean metric $\delta_{ab}$ where $a,b,c,...$ are spatial indices in the physical space. The generalized coordinates $\xi^i$ are indexed by $i,j,k$. With this Euclidean background metric, we introduce an induced metric that describes the distances between material elements as they move in time
\begin{align}
g_{ i j }  = \delta_{ab} ~\partial_{i }  X^a (\xi^k ,t ) \partial_j X^b (\xi^k ,t )~~.  
\end{align}  
The strain is then defined as the difference between this metric and a reference metric $g^{(0)}_{i  j}$ that describes the distances between material elements when the material is in an unstrained state, specifically
\begin{align}
\mathcal{E}_{i j }   & = \frac{1}{2} (  g_{i j } - g^{(0)}_{i  j} ) ~~.   \label{strainnn}
\end{align}
In order to track the evolution of the strain we define a covariant derivative of the form
\begin{align}
\frac{D}{D t }  A^{i ... m }_{\, \, \, \, j ... n }  = \dot{A}^{i ... m }_{\, \, \, \, j ... n } + \mathcal{L}_N A^{i ... m }_{\, \, \, \, j ... n }  \label{folpreservingggg}~~,
\end{align}
for any tensor $A^{i ... m }_{\, \, \, \, j ...n}$. $ \mathcal{L}_N$ is the Lie derivative along the vector $N^i$, it is given by:
\begin{align}
\begin{split}
\mathcal{L}_N  A^{i ...m }_{\, \, \, \, j ...n }  &  =   N^k \partial_k  A^{i ... m }_{\, \, \, \, j ... n } 
  \\  & - \partial_k N^i   A^{k ...m }_{\, \, \, \, j ... n }  - ...    - \partial_k N^m   A^{i ... k}_{\, \, \, \, j ... n }   \\  &  + \partial_j N^k   A^{i ... m }_{\, \, \, \, k ... n }   +   ...  + \partial_n N^k   A^{i ... m }_{\, \, \, \, j ... k }   ~~ . 
  \end{split}
\end{align} 
$N^i$  describes the movement of a fluid particle with respect to the coordinate (frame) choice in fluid space. For a change in time $\delta t $ the particle's coordinate $\xi^i$ moves with respect to the frame choice, i.e the coordinate of a particle initially located at $\xi^i$ changes to $\xi^i (t )  +  N^i (\xi^k ,t  ) \delta t $ in a time step $\delta t$ (see fig.~\ref{fehiufehiuh}).
Given the definition \eqref{folpreservingggg} we require the reference metric introduced in \eqref{strainnn} to be covariantly conserved
\begin{align}
 \frac{D}{D t } g^{(0)}_{i j } =0 \label{referencevanish}~~.
\end{align}
We thus view the reference metric just as an auxiliary metric that allows us to compare deformations at a given time $t$ with the original state of the material while simultaneously making a straightforward connection with the notion of strain commonly used in standard textbooks. Using \eqref{referencevanish} we obtain the following identity for the strain rate
\begin{align}
K_{i j } \equiv  \frac{D}{D t } \mathcal{E}_{i j } = \frac{1}{2}\frac{D}{D t } g_{i j }~~.  \label{fluidvelocityyyyy}
\end{align}
We will use these definitions to recast the model equations in a covariant form.

\subsection{Covariant model equations}
A description of the motion of a fluid can be obtained by attaching a set of coordinate labels  $X^a$ to each element of the fluid. The freedom in the choice of these labels corresponds to different frames \cite{Falkovich2011}. The velocity in any frame is given by \cite{Azeyanagi2009}:
\begin{align}
\frac{D}{D t}  X^a  = u^a  \label{fehiuehui}
\end{align}
And furthermore $u_i = u^a \partial_i X_a$. The introduction of the vector $N^i$ in \eqref{folpreservingggg} allows for frame choices that simplify our analysis. As we are working with infinitesimal strains, one possible choice, known as the comoving frame, is given by $X^a = X^a_{(0)}  +  w^a$ with $\dot{X}^a=u^a$ so that from eq.~\eqref{fehiuehui} it follows that $N^{i} = 0$. Here, $w^a$ is a small displacement vector and the scalar $X^{a}_{(0)}$ is defined such that $g_{ij}|_{X^a=X^{a}_{(0)}}=g_{ij}^{(0)}$. $u^i = u_{(0)}^i + \delta u^i$ is the fluid velocity, where $u_{(0)}^i$ denotes the velocity of the fluid in its initial state which obeys $\nabla_{i} u_{(0)}^{j} =0$ such that eq.~\eqref{referencevanish} is satisfied. Alternatively, one can work with the lab frame in which $N^i =u^i$ and  $X^a =\delta^{a  }_{\, \, i }  \xi^i$ such that $g_{ij} = \delta_{ij}$ and $\dot{X}^a = 0$. These two choices correspond to Lagrangian and Eulerian formulations of fluid dynamics. Either way we find
\begin{align}
    K_{i j }   =\frac{1}{2} \nabla_{(i} u_{j)}~~. \label{vielbeinnn}
\end{align}
The covariant derivative introduced in eq. \eqref{vielbeinnn} is defined in the usual way
\begin{align}
\begin{split}
\nabla_{i} A_j^{\, \, k } &  = \partial_{i} A_j^{\, \, k } - \Gamma^{l}_{i j } A_l^{\, \, k } + \Gamma^{k}_{i l } A_j^{\, \, l }~~,  \\ 
    \Gamma^{l}_{i j }    &   = \frac{1}{2}  g^{l k } ( \partial_{i } g_{k j}  +  \partial_{j } g_{k i} -  \partial_{k } g_{i j}  ) ~~,
\end{split}
\end{align}
for any tensor $A_j^{\, \, k }$ where $\Gamma^{l}_{i j } $ is the Christoffel connection build from the induced metric $g_{ij}$.
With these structures in hand, we now covariantize the definitions of sec. \ref{simplifiedddd}, in particular
\begin{align}
\eta_{i j k l }    & = \frac{1}{2} g_{ k (i} g_{j) l}  - \frac{1}{2} g_{i j } g_{kl }~~ \label{cov11},   \\ 
 \eta_{i j  k l  }^{*  }   &  =  -   \frac{1}{4}  (   \varepsilon_{i k } g_{j l}  + \varepsilon_{i l} g_{j k} + \varepsilon_{j k } g_{i l} +\varepsilon_{j l} g_{i k}    ) \label{cov12}~~,   \\ 
A_{\langle i j \rangle }  &   = \eta^{\, \, \, k l }_{i j } A_{k l  }  \, \, , \, \, 
A    = g^{i j  } A_{i j  }~~.
\end{align}
It is then straightforward to obtain the covariant form of the bulk \eqref{haaazzz} and shear \eqref{haahha} equations for the odd Jeffreys model which now take the form
\begin{align}
 \begin{split}
   &    \frac{D}{D t } \tau
   +  \sigma \tau    =   -  \alpha   \frac{D}{D t } K   -   \tilde\beta     K ~~,   \label{haaa}
 \end{split}  \\ 
 \begin{split}
   &   \frac{D}{D t } \tau_{\langle  i j  \rangle } 
   +  ( \chi \eta_{i j  }^{  \, \, \, \, \, k l }  + \chi^{*} \eta^{* \, \,  k l }_{i j }   )  \tau_{k l}   \\ &   =  -   ( \gamma \eta_{i j}^{  \, \, \, \, \, k l }  + \gamma^{*} \eta^{*  \, \, k l}_{i j  }   )    \frac{D}{D t } K_{kl}   -  ( \zeta \eta_{i j  }^{  \, \, \, \, \, k l }  + \zeta^{*} \eta^{* \, \, \,  k l }_{i j  }   )    K_{kl}    ~~.\label{haahha}
 \end{split}
\end{align}
In the following sections we choose the simple lab frame where $N^i =u^i$ and $g_{ij} = \delta_{ij}$ and we can thus use the partial derivative $\partial_i$ instead of the covariant derivative $\nabla_i$.

\subsection{Conservation laws}
\label{conservationlawsz}
The simplest way to embed Jeffreys model in a framework of viscoelasticity is to treat the stress $\tau_{ij}$ as dynamical degrees whose evolution is determined by \eqref{haaa}-\eqref{haahha}. Furthermore, one supplements the system with additional conservation laws that determine the evolution of hydrodynamic fields. In particular, the evolution of the fluid velocity $u^i$ introduced in \eqref{vielbeinnn}, the temperature $T$ and the mass chemical potential $\mu$ associated with mass density are respectively determined by
\begin{align}
      &  \rho  \dot{u}^{i} +   \rho u^{j}  \partial_j  u^{i}    + \partial_{j} t^{ i j  }     =   0~~,
 \label{momentumcons}      
 \\ 
 &      \dot{\epsilon} +    \partial_j ( \epsilon u^{j}  )+  \partial_j ( j^{j}_{\epsilon}  + u_{i} t^{ i j  } )    =   0~~,   \label{energycon} \\
 &  \dot{ \rho} +     \partial_j ( \rho u^{j}  )      =  0 \label{masscon} ~~.
\end{align}
Here $\epsilon$ and $\rho$ are the energy and mass density respectively, which are functions of $T, \mu$. We note that one can find different sign conventions for the stress in the literature. We furthermore defined 
\begin{align}
t^{i j }  &  = p  g^{i j }   + \tau^{i j }    \, \, , \, \,  \epsilon    = \epsilon_0    +\frac{1}{2} \rho u^2 ~~, 
\end{align}
where $p$ is the equilibrium pressure, $\epsilon_0$ is the part of the energy density that excludes the kinetic energy and $u^2=u_i u^i$. Both $p$ and $\epsilon_0$ are functions of $T$ and $\mu$. In addition, we introduced the heat current $j^{j}_{\epsilon}$, which contains gradient corrections that do not play a role in the analysis that we will carry out in this section. In sec.~\ref{sec:entropy} we show how we can explicitly derive the form of $j^{j}_{\epsilon}$. The conservation laws presented in \eqref{momentumcons}-\eqref{masscon} are typical of systems with Galilean symmetry.

\section{Entropy analysis} \label{sec:entropy}
In the previous sections we treated the stress $\tau_{ij}$ as dynamical degrees of freedom and stipulated its dynamics by means of phenomenological material diagrams. In this section we derive Jeffreys model from the second law of thermodynamics. In particular, we derive the constitutive relations, rheology equations and entropy constraints that arise from implementing the second law of thermodynamics \eqref{eq:2ndlaw} following the framework of refs. \cite{Fukuma_2011}. Using these general constitutive relations we demonstrate how to obtain the odd Jeffreys model that we introduced and studied in the previous sections. We also show how other simpler models (Kelvin-Voigt and Maxwell) arise as limiting cases of our general analysis, showing that parity-odd elastic terms are not compatible with entropy constraints, thus forbidding a passive incarnation of these models.

\subsection{Intrinsic metric}
Traditional hydrodynamic frameworks that implement the second law of thermodynamics in the context of viscoelastic fluids are based on introducing Goldstone fields of spontaneously broken translation symmetry \cite{PhysRevA.6.2401, chaikin2000principles, Armas2020}. However, as demonstrated in ref. \cite{Armas2020}, these formulations only allow for Kelvin-Voigt-type models and not, for instance, Jeffreys model. As such, we work with the formulation of refs. \cite{Azeyanagi2009, Fukuma_2011, Fukuma_2011_Entropic} in which metric degrees of freedom are introduced. 

Within this framework, a different notion of strain that also describes plastic deformations can be defined by introducing an evolving intrinsic metric $\tilde{g}_{ i j }(\xi^k,t)$ \cite{Fukuma_2011,Fukuma_2011_Entropic}. This metric measures the deformation of the distances between fluid particles as they are experienced due to bond configurations of the material. The difference between the induced metric and the intrinsic metric is a form of strain that is distinct from \ref{strainnn}, namely
\begin{align}
\kappa_{i j }  = \frac{1}{2} ( g_{ i j } -  \tilde{g}_{ i j } )  \label{strain}~~.
\end{align}
When a material is deformed and a strain is created, this intrinsic metric evolves as to cause \eqref{strain} to vanish. This can be understood as the bond configuration of the material, once the system is fully relaxed, adapting to the induced metric after the applied deformation. This change of the intrinsic metric is called a plastic deformation. The intrinsic metric $\tilde g_{ij}$ prior to any deformation is equal to the reference metric $g_{ij}^{(0)}$ introduced in \eqref{strainnn}.

\subsection{Thermodynamics and the second law}
\label{thermodynamicssss}
In this framework, the hydrodynamic fields consist of the fluid velocity $u^i$, temperature $T$, mass chemical potential $\mu$ and the intrinsic metric $\tilde g_{ij}$. The evolution of $u^i, T, \mu$ is again determined by the conservation laws \eqref{momentumcons}-\eqref{masscon} while the dynamics of $\tilde g_{ij}$ can be derived by supplementing the system with the second law of thermodynamics
\begin{equation} \label{eq:2ndlaw}
 \dot{s} +  \partial_j  (  s u^{j}  ) +  \partial_j  j_s^{j}     \equiv  \dot{s} +  \partial_j   s^{j}       =  \Delta~~,  
\end{equation}
where $s$ is the entropy density and $s^i$ is the entropy current given by $s^i = s u^i + j_s^{j}$. $j_s^{j}=j_\epsilon^j/T$ includes gradient corrections that will be determined\footnote{The identity $j_s^{j}=j_\epsilon^j/T$ follows from a convenient choice of hydrodynamic frame. A consequence of the same choice of hydrodynamic frame is the absence of gradient corrections to the mass conservation eq.~\eqref{masscon}.} and $\Delta$ is required to satisfy $\Delta\ge0$.
Local thermal equilibrium of our hydrodynamic theory naturally includes elasticity \cite{Armas2020}, which can be made explicit by considering the pressure 
\begin{align}
     p (  T, \mu  , \kappa_{ij}  )   =  p_f  (  T, \mu  )   - \frac{1}{2}  \lambda_1 \kappa^{\langle i j  \rangle }  \kappa_{\langle ij  \rangle }   - \frac{1}{4}  \lambda_2  \kappa^2   \label{pressuuuurere} ~~ ,
\end{align}
where $p_f  (  T, \mu   )$ is the fluid part of the equation of state and is independent of strain. The coefficients $\lambda_1$ and $\lambda_2$ are constrained to be positive due to mechanical stability \cite{Fukuma_2011} and denote the shear and bulk modulus respectively \cite{CallanJones2011}. To make this section more compact, we define:
\begin{align}
- \lambda_1 \kappa^{\langle i j  \rangle }  \kappa_{\langle ij  \rangle }   -  \frac{1}{2} \lambda_2  \kappa^2  \equiv   r^{ij} \kappa_{ij} ~~.  \label{simplifyy}
\end{align}
     From eq.~\eqref{pressuuuurere} it follows that in thermal equilibrium we have the following thermodynamic identities:
\begin{align} \label{eq:thermo}
\begin{split}
d\epsilon_0  & =Tds+\mu d\rho   - r^{ij} d \kappa_{ i j }  ~~, \\ 
d p   & =s d T +\rho  d \mu    +  r^{ij} d \kappa_{i j }  ~~, \\ 
~~\epsilon_0  & =- p  + Ts+\rho\mu~~,
\end{split}
\end{align}
characteristic of a system with Galilean symmetry. Instead of treating the stresses $\tau_{ij}$ as dynamical degrees of freedom, the second law \eqref{eq:2ndlaw} allows one to derive constitutive relations for $\tau_{ij}$ in terms of $u^i, T, \mu,\kappa_{ij}$ and rheological equations for the evolution of $\kappa_{ij}$. When combined, Jeffreys model is obtained as we will show in the next sections.

\subsection{Constraints from entropy production}
\label{constrheol}
We now focus on the second law of thermodynamics \eqref{eq:2ndlaw} and use it to derive constitutive relations and rheology equations. Using the conservation laws \eqref{momentumcons}-\eqref{masscon} together with the thermodynamic relations \eqref{eq:thermo} we explicitly evaluate the left hand side of \eqref{eq:2ndlaw} and obtain \cite{CallanJones2011}
\begin{align}
\begin{split}
\Delta  =  -  \tau^{ i j   }   \frac{ K_{ i j   }}{T}   -   \frac{1}{T^2}  j_{\epsilon}^i  \partial_i  T   & -\frac{  \lambda_1 }{ T}   \kappa^{\langle  i j  \rangle }  \frac{D}{D t }  \kappa_{\langle   i j \rangle}        -\frac{  \lambda_2 }{  2 T}    \kappa  \frac{D}{D t }  \kappa    ~. 
   \label{entropyyyyy}
\end{split}
\end{align}
All terms on the right hand side of eq.~\eqref{entropyyyyy} should be understood as being gradient suppressed. In particular,  we are assuming the following power counting scheme\footnote{Gradients of the mass chemical potential $\partial_i \mu$ are also of order $\mathcal{O} (\partial)$. However, they do not explicitly appear in our analysis due to the choice of hydrodynamic frame.}
\begin{align}
\kappa_{   i j } , \partial_{i } u_{ j }  , \partial_i T    = \mathcal{O} (\partial)  \label{powercounttting} ~~,
\end{align}
which implies that we are treating the strain, gradients of the fluid velocity \footnote{Eq.~\eqref{powercounttting} means that both vorticity, which is the anti-symmetrized fluid velocity gradient, as well as strain rate, which is the symmetrized velocity gradient, are first order.}, and temperature to be of the same relative order. 

Imposing $\Delta\ge0$ as required by the second law allows us to determine the constitutive relations for $\tau_{ij}$ and $j_\epsilon^i$ as well as the form of the rheology equations for the shear $\frac{D}{D t }  \kappa_{\langle   i j \rangle}$ and bulk $\frac{D}{D t }  \kappa$ sectors.\footnote{Note that while $\kappa_{i j}\sim \mathcal{O} (\partial)$, terms of the form $\kappa \frac{D}{Dt} \kappa$ are of order $ \mathcal{O} (\partial^2)$ \cite{Fukuma_2011}.} This analysis can be split into parity-even and parity-odd contributions as well as in scalar, vector and traceless tensor contributions to the constitutive relations. In the parity-even sector we obtain the following contributions\footnote{In both the parity-even and parity-odd sectors we have not included non-canonical terms besides the elastic terms. These potential additional terms correspond to hydrostatic transport as discussed in ref. \cite{jainbanerjeeeeeedutta} for the Galilean case and more extensively in refs. \cite{Jensen2012,Banerjee2012} for the relativistic case.} \footnote{It is possible to do a rescaling such that the coefficients $F$ and $G$, when positive, can be set to the coefficients $\lambda_1$ and $\lambda_2$ which allows one to arrive at the result in ref. \cite{CallanJones2011}.}
\begin{align}
\begin{split}
  j^{i}_{\epsilon}   =  &    -     \frac{L}{T^2 }  \partial^i  T ~~,
\end{split}
\\
\begin{split}
 \begin{pmatrix}      \tau_{\langle i j  \rangle}       \\  - \frac{  \lambda_1 }{ T}   \frac{D}{D t } \kappa_{\langle  i j   \rangle }    \end{pmatrix}   & =       \begin{pmatrix}  K_1  & -   G \\   G & K_2  \end{pmatrix}   \begin{pmatrix}    - \frac{1}{T}  K_{\langle i j  \rangle     }   \\   \kappa_{ \langle   i j  \rangle  }     \end{pmatrix} , \label{shearmatrix}
\end{split} \\ 
\begin{pmatrix}   \tau  \\   - \frac{ \lambda_2 }{ T}   \frac{D}{D t } \kappa     \end{pmatrix}    &  =   \begin{pmatrix}  M_1  &  -  F   \\   F & M_2   \end{pmatrix}    \begin{pmatrix}   - \frac{K}{T}   \\  \kappa \end{pmatrix} , \label{bulkmatrix}
\end{align}
where the coefficients $L, G, K_1,K_2,F,M_1,M_2$ are functions of $T$ and $\mu$. In order for the condition $\Delta\ge0$ to be satisfied, we need to impose
\begin{align} \label{firstconsss}
L,K_1,K_2,M_1,M_2 \ge0. 
 \end{align}
 Coefficients $F,G$ are arbitrary  and denote non-dissipative contributions to the constitutive relations \cite{Julicher_2018}. In turn, the parity-odd contributions take the form
\begin{align}
\begin{split}
   j^{i}_{\epsilon}   =  &     -   \frac{\beta}{T^2 } \varepsilon^{i}_j  \partial^j T ~~,  
\end{split}
\\ 
\begin{split}
\begin{pmatrix}      \tau_{\langle i j  \rangle}       \\  - \frac{  \lambda_1 }{ T}   \frac{D}{D t } \kappa_{\langle  i j   \rangle }    \end{pmatrix}   & =         
 \eta_{i j }^{* \,  \, \,  k l  }    \begin{pmatrix}    \alpha_1 & -  \alpha_3  \\   \alpha_3  &  \alpha_2  \end{pmatrix}      \begin{pmatrix}    - \frac{1}{T}  K_{ k l    }   \\   \kappa_{  k l }     \end{pmatrix}~~,  \label{odd22}
\end{split} 
\end{align}
where $\beta,\alpha_1,\alpha_2$ are arbitrary functions of $T$ and $\mu$ and characterise non-dissipative transport. In turn, $\alpha_3$ is an off-diagonal dissipative coefficient, function of $T$ and $\mu$, that in order to have $\Delta \geq 0 $ must satisfy
\begin{align}
\alpha^2_3 \leq    K_1 K_2 ~~.  \label{constraintttt}
\end{align}
Lastly, time-reversal invariance has been taken into account while defining the $K$, $M$ and $\alpha$ matrices \cite{CallanJones2011, Julicher_2018}, leading to the vanishing of the off-diagonal components in the case of the $K$ and $M$ matrices and a vanishing symmetric off-diagonal contribution to the $\alpha$ matrix.

\subsection{Viscoelastic models}
\label{viscoelasticmodelsss}
Given the constitutive relations and rheology equations derived in sec.~\ref{constrheol} we are in a position of deriving the models that we discussed throughout this paper. Using \eqref{shearmatrix}-\eqref{odd22}, we find the constitutive relations 
\begin{align}
\begin{split}
   \tau_{\langle i j \rangle }  &  =   -  (  K_1  \eta^{\, \,  \, \, \, k l }_{i j } 
 + \alpha_1  \eta^{* \, \, \, k l }_{i j }   )  \frac{1}{T} K_{k l } \\ &   -  (  G  \eta^{\, \,  \, \, \, k l }_{i j } + \alpha_3  \eta^{* \, \, \, k l }_{i j }   )  \kappa_{k l }  ~~,    \\ 
  \tau    &  =  -       \frac{ M_1}{T}  K     -  F  \kappa ~~,
\label{constt}
\end{split}
\end{align}
as well as the rheology equations 
\begin{align}
\begin{split}
  - \frac{ \lambda_1 }{ T}   \frac{D}{D t } \kappa_{\langle i j \rangle }  &  =   -  ( G   \eta^{\, \,  \, \, \, k l }_{i j } 
 + \alpha_3  \eta^{* \, \, \, k l }_{i j }   )  \frac{1}{T} K_{k l } \\ &   +   ( K_2 \eta^{\, \,  \, \, \, k l }_{i j } +  \alpha_2  \eta^{* \, \, \, k l }_{i j }   )  \kappa_{k l } ~~,     \\ 
 - \frac{ \lambda_2 }{ T}   \frac{D}{D t } \kappa   &  =  -        \frac{F}{T}  K  +    M_2  \kappa  ~~.
\label{rheologyyy}
\end{split}
\end{align}
This rheology equation describes the evolution of the strain \eqref{strain}, including the evolution of the intrinsic metric $\tilde g_{ij}$. The intrinsic metric is a dynamical degree of freedom which makes the system non-Markovian and therefore difficult to deal with. 

The most obvious way to simplify the set of eqs. \eqref{constt}-\eqref{rheologyyy} is by constraining the intrinsic metric to be non-dynamical, which corresponds to the elastic limit. This is achieved by requiring the intrinsic metric to satisfy
\begin{align}
 \frac{D}{D t }  \tilde{g}_{i j }   =0~~.\label{constrainttt}
\end{align}
In this context, $\tilde g_{ij}=g_{ij}^{(0)}$, where $g_{ij}^{(0)}$ was introduced in \eqref{strainnn}, and the notions of strain we introduced in sec.~\ref{strainnnn} coincide, namely $\mathcal{E}_{ij}=\kappa_{ij}$. The condition \eqref{constrainttt} fixes almost all coefficients appearing in \ref{rheologyyy}. In fact, using eqs.~ \eqref{folpreservingggg} and \eqref{strain} we find
\begin{align}
\begin{split}
     & K_2  =   \alpha_3    =  \alpha_2 = M_2   =0~~,  \\ 
   &  G  =      \lambda_1     \ \ , \ \ 
        F   =    \lambda_2 ~~.    \label{coefficientssss}
\end{split}
\end{align}
We note that \eqref{constrainttt} forces the coefficients $M_2,K_2$ to vanish. Imposing \ref{coefficientssss} in the constitutive relations \eqref{constt} leads to
\begin{align}
\begin{split}
   \tau_{\langle i j \rangle }  &  =   -  (  K_1  \eta^{\, \,  \, \, \, k l }_{i j } 
 + \alpha_1  \eta^{* \, \, \, k l }_{i j }   )  \frac{1}{T} K_{k l } -   \lambda_1 \eta^{\, \,  \, \, \, k l }_{i j }   \kappa_{k l } ~~,     \\ 
  \tau   &  =   -M_1  \frac{1}{T} K      -     \lambda_2 \kappa ~~.
\label{consttdddf}
\end{split}
\end{align}
Comparing this with the defining equation of the shear sector of the odd Kelvin-Voigt model \eqref{haaaz} we identify
\begin{equation}
\phi=\lambda_1~~,~~\phi^*=0~~,~~\psi=K_1~~,~~\psi^*=\alpha_1~~.
\end{equation}
As advertised in sec.~\ref{simplifiedddd}, the entropy constraints require that $\phi^*=0$ and so no odd elastic contributions to the constitutive relations for the stresses are allowed. The coefficient $\psi^*$ is allowed and characterises odd viscosity. As noted in ref. \cite{Scheibner_2020,Banerjee2021} a non-zero $\phi^*$ coefficient can be added to the model in the case of an active system but here we note that this cannot be the case in a passive one.

Having discussed this simpler case in which the condition \eqref{constrainttt} is enforced, we turn to the most general case. It is convenient to begin by focusing on the bulk sector of \eqref{constt} and \eqref{rheologyyy} involving $\tau$ and $\kappa$ as this sector does not contain parity-odd contributions. Acting with $D/Dt$ on the bulk sector of eq.~\eqref{constt} and using the bulk sector of \eqref{rheologyyy} leads to 
\begin{align}
\begin{split}
    \frac{D}{D t }   \tau    +    \frac{ M_2 T }{ \lambda_2 }   \tau   &  =-      \frac{M_1}{T}   \frac{D}{D t } K    +    \frac{ -   M_1 M_2 -F^2      }  { \lambda_2 }   K~~.    \label{finallll}
\end{split}
\end{align}
This equation corresponds to the bulk sector of Jeffreys model \cite{jeffrey1,jeffrey2,jeffrey3,jeffrey4,jeffrey5}. Indeed, comparing eq.~\eqref{finallll} with eq.~\eqref{haaazzz} leads to
\begin{equation}
\sigma=\frac{M_2 T}{\lambda_2}~,~\tilde\beta=\frac{M_1M_2 + F^2}{\lambda_2}~,~\alpha= \frac{M_1}{T}~~.  \label{bulkmodessssss}
\end{equation}
We now consider the shear sector of eqs. \eqref{constt} and \eqref{rheologyyy} which involves parity-odd terms. In order to find an equation of the form \eqref{haahha} it is useful to make use of the following identities
\begin{align}
     \eta_{i j k l  }  \eta^{* k l m n  }  &    =   \eta_{i j }^{ * \, \, \,  m n  } ~~,   \\ 
        \eta_{i j k l  }  \eta^{ k l m n  }  &    =   \eta_{i j  }^{ \, \, \, \, \,   m n  } ~~, \\ 
           \eta^{*}_{i j k l  }  \eta^{* k l m n  }  &    =   -  \eta_{i j  }^{ \, \, \, \, \,  m n  } ~~,
\end{align}
which allows one to find the equation
   \begin{align}
    \begin{split}
  \eta^{i j  }_{\, \, \, \,   m n  }     &    = 
 \frac{  C_1  \eta^{i j  }_{\, \, \, \, k l   }  -  C_2  \eta^{* i j  }_{\, \, \, \, \,  \, k l   }   }{  C_1^2 + C_2^2 }    (C_1    \eta^{k l }_{\, \, \, \, m n  }  + C_2  \eta^{* k l }_{\, \, \, \, \, m n  }    ) ~~, \label{inversionnnn}
\end{split}
    \end{align}
  which holds for any coefficient $C_1$ and $C_2$.
With this in mind, we follow the same steps as in the bulk sector, and eventually find
\begin{align}
 \begin{split}
    \frac{D}{D t } \tau_{\langle i j \rangle } &  =  \Big\{ -   G \eta_{ij  }^{\, \, \, \,   k l }   -  \alpha_3 \eta^{* \, \,  k l }_{ij  }   \Big\}  \\  &  \cdot   \Bigg[  - \Gamma   (  \Omega \eta_{k l  }^{\, \, \, \, \, m n } +  \Omega^{*} \eta^{* \, \, \, m n }_{k l   } )  \frac{1}{T}    \frac{D}{D t }  K_{m n }     \\ 
 &  -   \frac{T \Gamma}{ \lambda_1 }   (   K_2 \eta_{k l   }^{ \, \, \, \, \, m n  }  + \alpha_2 \eta^{* \, \, \,  m n }_{k l   }    )  \\ & \cdot  \Big\{    -   G    \eta_{ m n    }^{ \, \, \, \, \, \, o p  }  \tau_{o p} +  \alpha_3 \eta^{* \, \, \, o p }_{ m n    }  \tau_{o p}   \\  &  +   (  \Omega \eta_{m n   }^{\, \, \, \, \, o p  } +  \Omega^{*} \eta^{* \, \, \, o p  }_{m n    } ) \frac{1}{T}  K_{o p}     \Big\}  \\  &   + \frac{T}{\lambda_1}  \Big\{     G \eta_{k l  }^{\, \, \,  \, \, m n }  + \alpha_3 \eta^{* \, \, m n }_{k l   }    \Big\} \frac{1}{T}  K_{m n } \Bigg],    \label{bigequation}
 \end{split}
\end{align}
where we defined the coefficients
\begin{align}
\begin{split}
 \Gamma  &  = \frac{1}{  G^2 + \alpha_3^2  } ~~, \\ 
\Omega   &  = -G  K_1  - \alpha_3 \alpha_1 ~~,     \\ 
\Omega^{*}   &  =- \alpha_1    G + K_1  \alpha_3~~.   
\end{split}
\end{align}
Comparing \eqref{bigequation} with the characteristic equation of the shear sector of Jeffreys model \eqref{haahha} we identify the following coefficients
\begin{align}
\begin{split}
    \chi   &  =  \frac{  T }{ \lambda_1  } K_2   \, \, , \, \, 
      \chi^{*}     = \frac{  T}{ \lambda_1  }  \alpha_2 ~~,     \\ 
        \gamma   &  =   \frac{  K_1 }{T}    \, \, , \, \, 
      \gamma^{*}     =     \frac{ \alpha_1 }{T}~~,    \\ 
      \zeta   &  =   \frac{ 1}{ \lambda_1  }  (   -\alpha_1 \alpha_2-\alpha_3^2+G^2+K_1 K_2  )~~,  \\ 
      \zeta^{*}   &  =   \frac{ 1}{ \lambda_1  }  ( \alpha_2 K_1+\alpha_1 K_2 + 2 \alpha_3 G )~~.  \label{coefficientsss}
      \end{split}
\end{align}
This is the general form of Jeffreys model compatible with entropy constraints. An interesting limit of Jeffreys model with coefficients \eqref{coefficientsss} is obtained by setting $\alpha_1=K_1=0$ which, due to \eqref{constraintttt}, implies $\alpha_3=0$. This leads to $\gamma=\gamma^*=\zeta^*=0$. Comparing this case with the characteristic equation of the odd Maxwell model \eqref{haaahh} leads to
\begin{align}
\begin{split}
    \chi    =  \frac{  T }{ \lambda_1  } K_2   \, \, , \, \, 
      \chi^{*}     = \frac{  T}{ \lambda_1  }  \alpha_2       \, \, , \, \,    
      \zeta    =  \frac{ 1}{ \lambda_1  }  G^2      \, \, . \, \, 
     \label{maxwelll}
      \end{split}
\end{align}
Also, as advertised in sec.~\ref{simplifiedddd} the coefficient $\zeta^*=0$ vanishes for the odd Maxwell model due to entropy constraints.

\section{Modes}
\label{modess}
We are now in a position to study the collective modes corresponding to linear fluctuations of the two-dimensional parity-odd model presented in sec. s\ref{simplifiedddd}. We impose the entropy constraints of sec. \ref{sec:entropy} on the coefficients. For the calculation details we refer the reader to the appendix \ref{modecomutation}. We find the following shear and bulk damped modes due to the relaxation in the rheology equation
\begin{align}
    \omega_{1,2} &  = -i  ( \chi \pm i \chi^* )  -i k^2  \frac{\gamma \pm    i \gamma^*- \frac{\zeta  \pm i \zeta^*}{ \chi \pm  i \chi^*}}{ 2 \rho_{(0)}} + \mathcal{O} (k^3 ) \label{sheardamp}   ~~ , \\ 
       \omega_{3} &  = - i  \sigma    - i  k^2   \frac{-\tilde \beta  +\alpha  \sigma }{2 \sigma  \rho_{(0)}} + \mathcal{O} (k^3 )  \label{bulkkkdamp} ~~ .
\end{align}
In addition we observe the following sound modes and a diffusive mode
\begin{align}
\begin{split}
   \omega_{4,5}  & =   \pm k  \sqrt{\xi} - i k^2  \frac{     \zeta  \chi  + \zeta^*  \chi^{* }   }{4   \rho_{(0)} \left(\chi ^2+\chi^{* 2}\right)} \\  &    - i k^2  \frac{ \tilde \beta   }{4 \sigma  \rho_{(0)}  } + \mathcal{O} (k^3 )  \label{sounddd}  ~~ ,
   \end{split}  \\ 
   \omega_{6}   & =   - i k^2  \frac{   \zeta^* \chi^{* }+\zeta  \chi }{2 \rho_{(0)} \left(\chi ^2+\chi^{* 2}\right)} + \mathcal{O} (k^3 ) ~~ ,  \label{diffusivevvvv}
\end{align}
where $\xi$ is defined as a coefficient in the equation of state:
\begin{align}
p = p_0  +  \xi (\rho - \rho_{(0)})
\end{align}
On general grounds we expect systems satisfying the second law of thermodynamics to be stable.  The collective modes are stable if their diffusive and damping contributions have negative imaginary parts. In order to see this explicitly we need to impose the constraints coming form the second law of thermodynamics as given by eqs.~\eqref{bulkmodessssss} and \eqref{coefficientsss}. Due to the constraints on the signs of the transport coefficients the stability of the modes in eqs.~\eqref{sheardamp} and \eqref{bulkkkdamp} follows immediately. However, showing the stability of modes given by eqs.~\eqref{sounddd} and \eqref{diffusivevvvv} is more involved. We plug eqs.~\eqref{bulkmodessssss} and \eqref{coefficientsss} into the complex diffusive terms, which leads to the following condition:
\begin{align}
\begin{split}
 &   \zeta  \chi  + \zeta^*  \chi^{* }   \\  & =  \frac{T \left(G^2 K_2+2 \alpha_2 \alpha_3 G+K_1 \left(\alpha_2^2+K_2^2\right)-\alpha_3^2 K_2\right)}{\lambda_1^2}  ~~ .  \label{fenieniuniun}
 \end{split}
\end{align}
To show that eq. \eqref{fenieniuniun} must be non-negative we use the entropy constraint of eq.~\eqref{constraintttt}. This constraint allows us to define $ \alpha_3 = \pm ( \sqrt{K_1 K_2} - \upsilon )$, with $0 \leq \upsilon \leq  \sqrt{K_1 K_2}  $ and together with eq.~\eqref{firstconsss} we find
\begin{align}
\begin{split}
 &     \zeta  \chi  + \zeta^*  \chi^{* } =  \frac{T \left(G \sqrt{K_2}  \pm \alpha_2  \sqrt{ K_1 }\right)^2 }{\lambda_1^2}  \\  & + \upsilon  \frac{ T \left(K_2 \left( 2 \sqrt{K_1 K_2}-\upsilon \right) \mp 2  \alpha_2 G\right)}{\lambda_1^2} \\  &  \geq  T \frac{ \left(G \sqrt{K_2} \right)^2   \pm  2 (  \sqrt{K_1 K_2} - \upsilon  )  \alpha_2 G    + \left( \alpha_2  \sqrt{ K_1 }\right)^2 }{\lambda_1^2}  \\ &   + \upsilon  \frac{ T K_2  \sqrt{K_1 K_2} }{\lambda_1^2} \geq  0  ~~ .
   \end{split}
\end{align}
It follows that, for small wavenumbers, stability is guaranteed for all six modes. Lastly, we consider the limiting case $\upsilon=0$ leading to
\begin{align}
\begin{split}
 &    \zeta  \chi  + \zeta^*  \chi^{* }  =  \frac{T \left(G \sqrt{K_2}  \pm \alpha_2  \sqrt{ K_1 }\right)^2 }{\lambda_1^2}   ~~ .
   \end{split}
\end{align}
For $G \sqrt{K_2}  = \mp \alpha_2  \sqrt{ K_1 }$ we see that the modes have a vanishing imaginary part at the lowest order and thus the entropy constraints \eqref{constraintttt} still lead to stability.

\section{Discussion} \label{sec:discussion}
We have demonstrated that even though odd viscoelastic solids can only be active, odd viscoelastic fluids, which contain transient odd elasticity, can exist without an active driving. To achieve this we have discussed an extension of rheological Jeffreys model to chiral active media. We have furthermore shown that such responses leave clear imprints in the linear spectrum of fluctuations. Here, parity-odd elastic terms are analogous to transport coefficients such as Hall viscosity and Hall conductivity studied in the context of quantum matter. Our motivation, however, has been driven by the relevance of odd viscoelastic responses in biological systems and metamaterials.  

Metamaterials are artificially engineered structures, in which the properties of their constituents can be appropriately designed. One example of metamaterials consists of colloidal suspensions. It has been demonstrated that odd transport coefficients can be probed in a colloidal suspension of rotating particles suspended throughout a substance of larger molecules \cite{Soni2019}. Such active suspensions require a constant transfer of angular momentum provided by an external magnetic field, which presents an experimental challenge. Our analysis suggests that, since activity is not necessary to probe odd transport coefficients, a passive colloidal suspension of chiral objects such as granular particles \cite{Tsai2005} or helical nanoribbons \cite{Wang2013} is enough to see imprints of both odd viscosity and odd elasticity.

Parity-odd elastic responses are also relevant for chiral systems in quantum matter and high-energy physics in which the system may exhibit Lorentzian rather than Galilean symmetry. In fact, the method by which we first obtained some of the results presented in this paper was to first consider parity-odd responses in a higher-dimensional relativistic theory and later dimensionally reduce to arrive at a hydrodynamic theory with Galilean symmetry as in ref. \cite{jainbanerjeeeeeedutta}. These details will be given in another publication.

\section{Acknowledgements} 
JA is partly supported by the Nederlandse Organisatie voor Wetenschappelijk Onderzoek (NWO) through the NWA Startimpuls funding scheme and by the Dutch Institute for Emergent Phenomena (DIEP) cluster at the University of Amsterdam. RL was supported, in part, by the cluster of excellence ct.qmat (EXC 2147, project-id 39085490). PS acknowledges the support of the Narodowe Centrum Nauki (NCN) Sonata Bis grant 2019/34/E/ST3/00405 and NWO Klein grant via NWA route 2.
      
\appendix

\section{Derivation of viscoelastic dissipation rate}
\label{entropcurrentdissipationrate}
In this appendix we derive the dissipation rate presented in eq.~\eqref{entropyyyyy}. We start by rewriting the thermodynamical identities presented in sec. \ref{thermodynamicssss}. Because we work with Galilean symmetry, the particle number density $n$ can be related to momentum density $\rho$ via the particle mass $m$, i.e. $\rho = m n $, turning the momentum density into mass density.  Additionally we can absorb the velocity dependence into the mass chemical potential $\mu$ by taking $\mu = \nu /m+ \frac{1}{2}   u_i^2 $. Here, $\nu$ is an ordinary chemical potential that couples to $n$. We will now rewrite the thermodynamic identities from sec. \ref{conservationlawsz} so that the velocity dependence is explicitly shown, that is
\begin{align}
d  p  &  = s  d T   +  \pi_i  d u^i  + n d \nu  +  r^{ij}  d  \kappa_{ij}  \label{gibssdiheemmdw1z22}  \\ 
d \epsilon  &  = T d  s +  u_i d  \pi^i   + \nu d n    - r^{ij}  d  \kappa_{ij}   \label{firstlaww23z2}  \\ 
\epsilon &  = T s  - p  +  u_i \pi^i  + \nu n  \label{eulzew}      ~~.
\end{align}
Here we defined the momentum $\pi^i = \rho u^i $. Note that these thermodynamic identities are not written in terms of $\epsilon_0$ but in terms of $\epsilon = \epsilon_0 +  \frac{1}{2} u_i \pi^i$. We also rewrite the conservation laws in sec. \ref{conservationlawsz} as follows:
      \begin{align}
      &    \dot{\pi}^{i} +   \partial_j ( \pi^{j}    u^{i} )   + \partial_{j} t^{ i j  }     =   0~~,
 \label{momentumconsA}      
 \\ 
 &      \dot{\epsilon} +    \partial_j ( \epsilon u^{j}  )+  \partial_j ( j^{j}_{\epsilon}  + u_{i} t^{ i j  } )    =   0~~,   \label{energyconA} \\
 &  \dot{ n} +     \partial_j ( n u^{j}  )      =  0 \label{massconA} ~~ ,  \\ 
   & \rho  =   m n  ~~  .
\end{align}
   Using the first law in eq. \eqref{firstlaww23z2} as well as the conservation laws we find the rate of entropy equation
       \begin{align}
       \begin{split}
           \dot{s} & = \frac{1}{T}  \Big\{ \dot{\epsilon}  - u_i  \dot{\pi}^i   - \omega \dot{n} +  r^{ij} \dot{ \kappa}_{ij}      \Big\}  \\ 
    & =  \frac{1}{T}  \Big\{ - \partial_{i} ( \epsilon   u^i  + t^{ij} u_j + j^i_{(\epsilon)} )    + \omega   \partial_j ( n u^{j}  )    \\  & +  u_i  \partial_j ( \pi^i u^j)   + u_i  \partial_j ( p  g^{ij}  + \tau^{ij})   + r^{ij} \dot{ \kappa}_{ij}     \Big\} \\ 
    & = -\partial_i s^i     +\frac{1}{T}  \Big\{ \partial_{i}  [   u^i ( - \epsilon - p  +T s + \nu n  + \pi_i u^i   )] \\  & +  u^j (   -   \pi^i  \partial_j u_i  -s  \partial_{j} T +   \partial_{j} p - n \partial_j \nu  )   - \tau^{ij}  \partial_i  u_j   \\  &  + T  j^i_{(\epsilon)}   \partial_i( \frac{1}{T})    + r^{ij} \dot{ \kappa}_{ij}    \Big\}   \label{dissippp} ~~ .
    \end{split}
    \end{align}
    Using eq.~\eqref{gibssdiheemmdw1z22} and \eqref{eulzew} as well as eq. \eqref{eq:2ndlaw} we obtain
    \begin{align}
        \begin{split}
      T \Delta   & =  - \tau^{ij}  \partial_i  u_j  +  T  j^i_{(\epsilon)}   \partial_i( \frac{1}{T}) + r^{ij}   (  \dot{\kappa}_{ij}  +  u^k \partial_k  \kappa_{ij}    )  ~~ .
           \end{split}
       \end{align}
Due to the power counting introduced in eq. \eqref{powercounttting}, we can use the following perturbative identity \cite{Fukuma_2011,Azeyanagi2009} \begin{align}
  \dot{\kappa}_{ij}  +  u^k \partial_k  \kappa_{ij}       = \frac{D}{D t } \kappa_{ij}  + \mathcal{O} (\partial^2 )  \label{perttttt} ~~ ,
\end{align}
to simplify the results. This identity allows us to take the elastic limit in eq.~\eqref{constrainttt}. A non-perturbative covariant study of strain has been performed \cite{Armas2020} and will be extended to plastic deformations in a future work. Considering eqs. \eqref{perttttt}, \eqref{dissippp} and \eqref{dissippp} we derive eq.~\eqref{entropyyyyy}.

\section{Diagrams}
\label{app:diagrams}
In this appendix we consider a diagrammatic representation of the models that we introduced in sec. \ref{simplifiedddd}. This can be achieved by working with diagrams of the type presented in fig.~\ref{bulksectorsnip} in terms of electric circuits. Within this setting, the stress plays the role of the electric current, i.e. when drawing components in series, the stress going through the components is the same for every component. Similarly, the stress is divided when the components are connected in parallel. Several components can be considered in such circuits but the ones that we focus on are the spring and the dashpot. A spring plays the role akin to a resistance, which gives an amount of stress that is proportional to the strain, and the strain thus plays the role of potential. The dashpot also gives stress, but now it is induced by the time-derivative of the strain.  

To begin with we consider the circuit that we introduced in fig.~\ref{bulksectorsnip} which represents the bulk sector of Jeffreys model. Following the circuit rules the characteristic equations are
\begin{align} \label{eq:circuit1}
\begin{split}
    \tau  &  =   \tau_{(1)} + \tau_{(2)}~~,~~\mathcal{E}  = \mathcal{E}_{(1)} + \mathcal{E}_{(2)}~~,   \\ 
\tau_{(1)}  &  =  - A   \mathcal{E}_{(1)}~~,~~\tau_{(1)}  =  - B  K_{(2)} ~~,~~\tau_{(2)} =   - C  K~~,
\end{split}
\end{align}
where $A,B,C$ are arbitrary coefficients associated with each component in the diagram of fig.~\ref{bulksectorsnip}. We note that while the total stress $\tau$ and the total strain $\mathcal{E}$ are physical quantities, the interpretation of the individual stresses $\tau_{(1)},\tau_{(2)}$ and individual strains $\mathcal{E}_{(1)},\mathcal{E}_{(2)}$ is not always clear and such quantities should be thought of as auxiliary quantities. Given \eqref{eq:circuit1} we act with $D/Dt$ on $\tau_{(1)}$ and use the remaining identities to find  
\begin{align}
\begin{split}
  \frac{D}{D t } \tau +  \frac{A}{B}   \tau  &  =    -  C \frac{D}{D t} K     -  \left(A + \frac{A C }{B}  \right) K~~,   \end{split}
\end{align}
which when compared with eq.~\eqref{haaa} we identify
\begin{equation}
\sigma=\frac{A}{B}~~,~~\alpha=C~,~~\tilde\beta=A + \frac{A C }{B} ~~.
\end{equation}
\begin{figure}[ht]
\includegraphics[width=8cm]{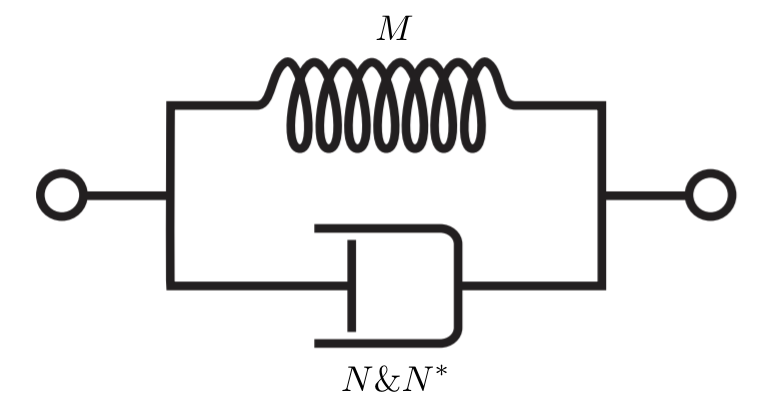}
\caption{Material diagram corresponding to the passive odd Kelvin-Voigt model.}
\label{parallelKV}
\end{figure}
Before addressing the shear sector of Jeffreys model, we consider the simpler case of the passive Kelvin-Voigt model represented in fig.~\ref{parallelKV}, from which we can immediately extract the constitutive equation
\begin{align}
    \tau_{\langle i j \rangle }  &  =    -  ( N  \eta^{\, \, \, \,  k l }_{i j }  + N^{*}  \eta^{* \, \,  k l }_{i j }) K_{k l }   -  M \eta^{\, \,  \, \, k l }_{i j }  \mathcal{E}_{k l } ~~.
\end{align}
Comparing this constitutive equation with eq.~\eqref{haaaz} we identify
\begin{equation}
\psi=N~~,~~\psi^*=N^*~~,~~\phi=M~~,~~\phi^*=0~~.
\end{equation}
Moving on to the shear sector of Jeffreys model represented in fig.~\ref{shearsectorsnip}, the circuit equations take the more intricate form
\begin{align}
\begin{split}
    \mathcal{E}_{\langle i j \rangle }  &  =  \mathcal{E}^{(1)}_{\langle i j \rangle } +   \mathcal{E}^{(2)}_{\langle i j \rangle }~~,   \\ 
        \tau_{\langle i j \rangle }  &  =   \tau^{(1)}_{\langle i j \rangle } +    \tau^{(2)}_{\langle i j \rangle } ~~,  \\ 
 \tau^{(1)}_{\langle i j \rangle }  &  =  -  (   U \eta_{i j  }^{\, \, \, \, \, k l}   + U^{*} \eta^{* \, \, k l }_{i j } )   \mathcal{E}^{(1)}_{kl} ~~,  \\ 
\tau^{(1)}_{\langle i j \rangle }  &  =  - (   V \eta_{i j  }^{\, \, \, \, \, k l}   + V^{*} \eta^{* \, \, k l }_{i j } )   K^{(2)}_{kl}~~,    \\ 
  \tau^{(2)}_{\langle i j \rangle }  &  =  -  (   W \eta_{i j }^{\, \, \, \, \, k l}   + W^{*} \eta^{* \, \, k l }_{i j  } ) K_{k l }~~.   \\ 
\end{split} \label{firstlinhheeeee}
\end{align}
As in the bulk sector, we can manipulate \eqref{firstlinhheeeee} in order to find the rheology equation
\begin{align}
\begin{split}
   &      \frac{D}{D t } \tau_{\langle i j \rangle }      =   -  (   W \eta_{i j}^{\, \, \, \, \, k l}   + W^{*} \eta^{* \, \,  k l}_{i j  } ) \frac{D}{D t} K_{k l }   \\   &  -   (   U \eta_{i j   }^{\, \, \, \, \, k l}   + U^{*} \eta^{* \, \, k l }_{i j} )   \big[K_{kl}   +     \Theta (   V \eta_{ k l   }^{\, \, \, \, \, m n }   - V^{*} \eta^{*\, \,  m n }_{ k l} ) \\  & \cdot ( \tau_{m n  }  +     W \eta_{  m n }^{\, \, \, \, \, o p } K_{o p  }   + W^{*} \eta^{* \, \, o p }_{ m n  }  K_{o p  } ) \big]~~,  \label{subsieuhlll}   \end{split} \end{align}
where, for convenience, we have defined 
\begin{align}
       \Theta = \frac{1}{ V^2  + V^{* 2 } }~~.
\end{align}
Comparing eq.~\eqref{subsieuhlll} with \eqref{haahha} one readily identifies
  \begin{align}
  \begin{split}
 \chi  & =      \Theta ( U V + U^{*} V^{*}    )  \, \, , \, \, 
 \chi^{*}   =      \Theta (  -   U V^{*} + U^{*} V  )~~,   \\ 
\gamma  & = W  \, \, , \, \, 
 \gamma^{*}   = W^*~~,  \\ 
  \zeta & = U+       \Theta ( U V  W - U^{*} V W^{*}+ U^{*} V^{*} W + U V^{*} W^{*} )~, \\ 
 \zeta^{*}  & = U^*  +      \Theta ( U^{*} V W + U V W^{*}- U V^{*} W + U^{*} V^{*} W^{*}  )~.
 \label{endresultt}     \end{split}
\end{align}
\begin{figure}[ht]
\includegraphics[width=8cm]{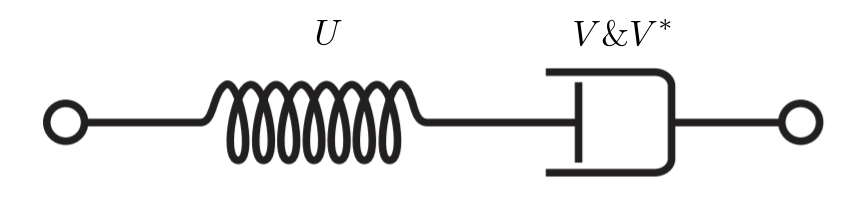}
\caption{Diagram corresponding to the passive odd Maxwell model. The \quotes{\&} refers to a parallel connection of an odd and an even component of the same type.}
\label{seriesmaxwell}
\end{figure}
It is also possible to invert the relations in \eqref{endresultt} and although we do not do this here explicitly as it is a very heavy operation, we have verified numerically that there are always real solutions. It is interesting to note that eq.~ \eqref{endresultt} implies that $\chi^{*}$ can only be non-zero if both even and odd components are non-vanishing, whereas all other components can be non-zero if only odd or only even components are non-zero. 

A limiting case of this model can be attained if we take $W,W^*,U^*$ to be zero. In this case we obtain the diagram given in fig.~\ref{seriesmaxwell}. This diagram corresponds to the passive odd Maxwell model introduced in eq.~\eqref{haaahh} with the following coefficients
 \begin{align}
  \begin{split}
 \chi  & =  \frac{ U  V}{V^2 + V^{* 2 }}   \, \, , \, \,  \chi^{*}   =  - \frac{  U  V^{*}  }{V^2 + V^{* 2 } }  \, \, , \, \, 
\gamma   = 0~~,  \\   \gamma^{*}  &  = 0  \, \, , \, \,   \zeta  = U   \, \, ,  \, \,  \zeta^{*}   =  0 ~~.
 \label{endresererultt}     \end{split}
\end{align}

\section{Computation of the modes}
\label{modecomutation}
In this appendix we give computational details about the modes of sec. \ref{modess}. The linear perturbations we consider are $\delta \rho$, $\delta u_i$ and $\delta \tau_{ij}$, i.e. we consider mass, fluid velocity and stress fluctuations. For simplicity we do not consider energy density fluctuations and therefore will not require working with the energy conservation equation. We use eqs. \eqref{momentumcons} and \eqref{masscon} as well as eqs.~\eqref{haaaz}, \eqref{haaahh} to get the following fluctuation equations
\begin{align}
\begin{split}
   \partial_t \delta  \tau_{\langle ij \rangle }   &  =- \chi  \eta_{i j k l } \delta \tau_{k l } -  \chi^* \eta^*_{i j k l } \delta \tau_{k l }  -  \gamma \eta_{i j k l } \partial_t \delta K_{k l }  \\  &  - \gamma^* \eta^*_{i j k l } \delta K_{k l }      - \zeta \eta_{i j k l } \delta K_{k l }  - \zeta^* \eta^*_{i j k l } \delta  K_{k l }  ~~ ,  \end{split} \\ 
   \partial_t \delta  \tau      &  = - \sigma    \delta \tau    - \alpha  \partial_t \delta K   -   \tilde \beta \delta  K ~~ ,  \\ 
                           \partial_t \delta \rho     & =  - \rho_{(0)} \partial_i \delta u_i  \label{densityu1}     ~~ ,  \\ 
                           \begin{split}
     \rho_{(0)}     \partial_t \delta u_{i }   & =   - \xi  \partial_i \delta \rho      -  \partial^{j} \delta \tau_{  i  j }  ~~ . 
     \end{split}
   \end{align}
Since we have a rotationally invariant system we are free to fix the spatial dependence of the fluctuations to be in the $x$-direction without loss of generality. We can write these linearized equations in a compact form as
\begin{align}
  ( Z_1  + Z_2 \partial_t   + Z_3 \partial_x    )  v   =0   \label{smallequation} ~~ .
\end{align}
The vector and matrices in eq. \eqref{smallequation} are given by
\begin{align}
\begin{split}
  v   &  = \begin{pmatrix} 
   \delta   \tau_{\langle x x  \rangle }  \\ 
      \delta   \tau_{ x y  } \\ 
     \frac{1}{2}    \delta         \tau \\ 
                     \delta \rho \\ 
               \delta u_x \\ 
                  \delta u_y \\ 
    \end{pmatrix}    \, \, , \, \, 
  Z_1    =   \begin{pmatrix}
 \chi    & -\chi^*  & 0   &  0 & 0  & 0     \\ 
 \chi^*  & \chi   & 0   &  0 & 0  & 0   \\ 
0  & 0  &  \sigma  &  0 & 0  & 0    \\ 
0 & 0  & 0  &0 & 0  & 0    \\ 
0 & 0  & 0  &0 & 0  & 0   \\ 
0 & 0  & 0  &0 & 0  & 0  
\end{pmatrix} ~~ , 
\end{split}  
\\ 
 Z_2  &         =       \text{diag} (1 , 1,1 , 1 ,\rho_{(0)}   , \rho_{(0)} ) ~~ ,
  \\   \begin{split}  
    Z_3      &  =     \begin{pmatrix}
  0 & 0  & 0 & 0 &   \frac{1}{2} (  \zeta  + \gamma \partial_t )  &  - \frac{1}{2} ( \zeta^*  + \gamma^*  \partial_t )  \\ 
  0 & 0 &  0 & 0 &  \frac{1}{2} (  \zeta^* + \gamma^*  \partial_t) &   \frac{1}{2} (\zeta + \gamma \partial_t )    \\ 
  0 & 0 & 0 & 0 & \frac{1}{2} ( \tilde \beta   + \alpha \partial_t   )    & 0   \\ 
  0 & 0 & 0 & 0 &  \rho_{(0)} &   0   \\ 
  1  & 0  & 1  &  \xi &  0 & 0    \\ 
0  & 1  &  0  &  0   & 0  & 0 
\end{pmatrix} ~~ . 
\end{split}  
\end{align}
The dispersion is found by considering plane waves of the form $\sim e^{- i \omega t  + i k x }$ and imposing $\text{det}(M)=0$ with
\begin{align}
 M  =       Z_1  +  (- i \omega )  Z_2   + i k  Z_3  (k , \omega )   ~~ . 
\end{align}
We then need to solve the following equation
\begin{align}
  \det (M)  =  (- i  \omega )^6   \rho^2_{(0)  }   + ....  =0  ~~ ,
\end{align}  
which has six solutions for $\omega$. They are presented up to the second order in $k$ in sec.~\ref{modess}.

\bibliographystyle{apsrev4-1}
\bibliography{passive_odd_elasticity}

\end{document}